\documentclass[sigplan,10pt]{acmart}
\usepackage{subcaption}
\usepackage{listings}
\usepackage{array}
\usepackage{calc}
\usepackage{xspace}
\usepackage{booktabs}
\usepackage[T1]{fontenc}
\usepackage{natbib}

\usepackage{tikz}
\usepackage{pifont}
\usetikzlibrary{arrows.meta, positioning, fit, decorations.pathreplacing,
                   backgrounds, shadows.blur}
\newcommand\systemname{\textsc{Meshlib}\xspace}
\author{Habib Mostafaei}
\orcid{0000-0001-8282-1571}
\affiliation{%
  \institution{Eindhoven University of Technology}
  \city{}
  \country{}
  }
\email{h.mostafaei@tue.nl}

\author{Tom Van Liempd}
\affiliation{%
  \institution{Eindhoven University of Technology}
  \city{}
  \country{}
}
\email{tom@berlehof.nl}
\renewcommand\footnotetextcopyrightpermission[1]{}
\acmConference[Submitted for review ]{}
\usepackage{listings}
\usepackage{xcolor}

\definecolor{loginFill}  {RGB}{255,235,210}
\definecolor{loginBorder}{RGB}{215,155,  0}
\definecolor{userFill}   {RGB}{210,220,240}
\definecolor{userBorder} {RGB}{100,120,180}
\definecolor{epBlueFill} {RGB}{218,232,252}
\definecolor{epBlueBdr}  {RGB}{108,142,191}
\definecolor{epPinkFill} {RGB}{235,215,235}
\definecolor{epPinkBdr}  {RGB}{170,130,180}
\definecolor{lstclr@kw}     {RGB}{  0, 70,160}   
\definecolor{lstclr@type}   {RGB}{ 80,  0,140}   
\definecolor{lstclr@comment}{RGB}{100,100,100}   
\definecolor{lstclr@string} {RGB}{160, 32, 32}   
\definecolor{lstclr@num}    {RGB}{ 80,120, 80}   
\definecolor{lstclr@rule}   {RGB}{180,180,180}   

\lstdefinestyle{macroTemplate}{
  basicstyle      = \ttfamily\fontsize{7.5}{9}\selectfont,
  keywordstyle    = \bfseries\color{lstclr@kw},
  keywordstyle    = [2]\color{lstclr@type},       
  keywordstyle    = [3]\color{lstclr@num},        
  commentstyle    = \itshape\color{lstclr@comment},
  stringstyle     = \color{lstclr@string},
  frame           = l,
  framerule       = 0.6pt,
  rulecolor       = \color{lstclr@rule},
  framesep        = 3pt,
  xleftmargin     = 6pt,
  xrightmargin    = 0pt,
  backgroundcolor = {},
  columns         = fullflexible,
  keepspaces      = true,
  tabsize         = 2,
  showstringspaces= false,
  numbers         = none,
  breaklines      = true,
  breakatwhitespace = true,
  breakindent     = 1em,
  captionpos      = t,
  aboveskip       = 4pt,
  belowskip       = 2pt,
  abovecaptionskip= 0pt,
  belowcaptionskip= 1pt,
  lineskip        = 0pt,
}

\lstdefinelanguage{protobuf}{
  morekeywords    = [1]{syntax, package, option, service, rpc,
                        returns, stream, message, oneof, enum,
                        repeated, optional, reserved},
  morekeywords    = [2]{bool, uint32, uint64, int32, int64,
                        bytes, string, float, double},
  morekeywords    = [3]{true, false, 0, 1, 2, 3, 4},
  morecomment     = [l]{//},
  morecomment     = [s]{/*}{*/},
  morestring      = [b]",
  sensitive       = true,
}
 
\lstdefinelanguage{YAML}{
  morekeywords    = [1]{apiVersion, kind, metadata, spec,
                        endpointSelector, matchLabels,
                        ingress, fromEndpoints, toPorts,
                        ports, protocol, rules, http,
                        name, app, port, method, path},
  morekeywords    = [2]{true, false, null},
  morecomment     = [l]{\#},
  morestring      = [b]",
  morestring      = [b]',
  sensitive       = true,
}

\usepackage{float}          
\newfloat{codelisting}{t}{lol}
\floatname{codelisting}{Listing}
 
 \newcommand{\cross}{$\times$}
\newcommand{\circnum}[1]{%
\tikz[baseline=(char.base)]{
\node[
    shape=circle,
    fill=black,
    text=white,
    draw=black,
    inner sep=0.8pt
] (char) {\normalsize #1};}}

\begin{document}

\title{\systemname: In-Process Policy Enforcement for Sidecar-less Service Meshes}

\begin{abstract}
Service meshes facilitate service-to-service communication and enforce security policies in microservice architectures. However, they often depend on per-pod sidecar proxies, which introduce significant latency and resource overhead due to redundant application-layer parsing on every request. Eliminating sidecars without compromising security guarantees remains a central challenge.
To address this, we introduce \systemname,  a sidecar-less
service mesh extension built on Cilium as a control-plane extension. \systemname incorporates a non-intrusive application-bound library that enforces Layer-7 policies within the application process, while delegating transport-level identity and routing to Cilium's eBPF-based data plane. This separation of responsibilities removes sidecar-induced latency and maintains the security semantics of the service mesh. The architecture remains fully interoperable with unmodified services, enabling incremental adoption within existing deployments. 
We evaluate \systemname{} against Istio, Linkerd, and unmodified Cilium on the TrainTicket benchmark, enforcing 126 security policies across 37 services and show that it achieves the lowest end-to-end latency of all evaluated configurations with comparable resource overhead.

\end{abstract}

\maketitle 
\pagestyle{plain}

\section{Introduction}\label{sec:intro}

Modern cloud-native applications are increasingly built as collections of loosely coupled microservices, each independently deployed and communicating over a network~\cite{gan_open-source_2019,kubernetes,evoluationOfMicroservices}. As deployments grow, managing service-to-service communication, including security policy enforcement~\cite{ServiceRouter-OSDI23,ResourceAllocationMeta-OSDI24,safeTree-OOPSLA25}, access control~\cite{automaticPolicy-Security21,AccessWeb-SP25,accessControl-CCS25,RBAC}, and traffic routing~\cite{Rajomon-NSDI25,michaelis_l3_2024,HardMesh-SIGCOMM25,MeshTest-NSDI25}, becomes a significant operational challenge, which has driven the widespread adoption of service meshes~\cite{saxena_invited_2023,michaelis_l3_2024,ServiceRouter-OSDI23}. Service meshes enforce these policies by deploying a dedicated proxy, a \emph{sidecar}, alongside each service instance, intercepting all inbound and outbound traffic transparently~\cite{zhu_dissecting_2023}. They enable developers to focus on implementing business logic rather than managing infrastructure complexities~\cite{michaelis_l3_2024}.

In such microservice deployments, all inter-service communication is performed over the network, inherently expanding the attack surface. In particular, a compromised microservice can act as a pivot point, issuing malicious requests to other services and enabling lateral movement across the application~\cite{accessControl-CCS25,automaticPolicy-Security21}. The implicit trust exacerbates this threat, often assumed within cluster-internal communication.

To mitigate these risks, modern orchestration frameworks, such as Kubernetes~\cite{kubernetes,evoluationOfMicroservices} and service meshes such as Istio~\cite{istio} and Cilium~\cite{noauthor_cilium_nodate}, provide fine-grained access control mechanisms between services. These mechanisms allow operators to define and enforce policies that restrict which services can communicate and under what conditions, forming a critical component of the security posture in microservice-based systems~\cite{automaticPolicy-Security21,Mazu-EuroSec25,KubeSecurity-IEEESP21,eZTrust-SOSR19}.

Enforcing such policies, however, exposes a fundamental trade-off in current service mesh designs. Application-layer (Layer~7) enforcement enables expressive policies, such as endpoint-level access control based on HTTP paths and methods, but requires intercepting and parsing every request~\cite{automaticPolicy-Security21,expressiveness-HotNets23}. Existing service meshes intercept all inter-service traffic through a dedicated sidecar or proxy, which must fully parse each request to evaluate policies before forwarding it.
This redundant parsing by the application layer (the request is parsed again by the receiving application) increases end-to-end latency by approximately 3$\times$ in Cilium~\cite{zhu_dissecting_2023}.
At scale, these per-request costs accumulate along every hop of a multi-service call path and translate directly into infrastructure cost and degraded user experience~\cite{ServiceRouter-OSDI23,dean_tail_2013}. In contrast, transport-layer (Layer 4) enforcement avoids most of this overhead, but can express only coarse-grained policies at the level of entire services, which is insufficient for many security requirements. As a result, developers face a difficult choice between precise, application-aware policy enforcement and low-latency communication. This tension raises a key question: \textit{Can Layer~7 policy enforcement be decoupled from proxy-based traffic interception without sacrificing security guarantees, deployability, or compatibility with unmodified services?}

We present \systemname{}, a sidecar-less service mesh extension for Cilium that enforces Layer~7  policies in-process within an optional application-bound library, while retaining full compatibility with services that do not adopt the library. \systemname{} coordinates enforcement state with the service mesh control plane to ensure that security guarantees are preserved during policy updates and incremental deployment.  \systemname{} reuses protocol structures already parsed by the application framework, eliminating the redundant marshal--inspect--unmarshal cycle that dominates sidecar latency. Services without the library fall back transparently to Cilium's existing Layer~7 Envoy proxy, enabling incremental
adoption without disrupting existing deployments. \systemname{} is
built on standard Kubernetes infrastructure and requires no changes to
unmodified services.

\noindent\textbf{Contributions.}
This paper makes the following contributions:
\begin{itemize}

  \item \textbf{Sidecar-less Layer~7 enforcement.}
        \systemname{} performs HTTP policy enforcement in-process via an optional library while relying on Cilium's eBPF data plane for transport identity and routing.

  \item \textbf{Consistent policy coordination.}
        We design a lightweight bidirectional gRPC protocol that activates in-process enforcement only after policy acknowledgment, eliminating update gaps and supporting transparent fallback for unmodified services.

  \item \textbf{Evaluation.}
        We evaluate \systemname{} against Istio, Linkerd,
        and unmodified Cilium on the TrainTicket
        benchmark~\cite{zhou_benchmarking_2018}, enforcing
        126 policies across 37 services; \systemname{}
        achieves the lowest end-to-end latency with
        comparable resource overhead.
 
  \item \textbf{Open-source artifact.}
        We release the full implementation, gRPC protocol
        definition, policy-generation tooling, and
        evaluation scripts as open-source software.

\end{itemize}

\section{Background and Motivation}\label{sec:motivation}

\subsection{Background}
\medskip
\noindent\textbf{Service meshes.} 
A \textit{service mesh} addresses the complexity of inter-service communication by abstracting it into a dedicated infrastructure layer. It comprises a data plane, which is a set of proxies that intercept and forward traffic between services, and a control plane that distributes configuration and policies to these proxies~\cite{saxena_invited_2023,HybridMesh-NSDI26}. In the prevalent sidecar model, a dedicated proxy container, typically Envoy, is deployed alongside each service instance. All inbound and outbound traffic is routed through this sidecar, which enforces policies configured by the control plane before forwarding requests to the application~\cite{survey,survey-2021}.

\noindent\textbf{eBPF.}
The Extended Berkeley Packet Filter (eBPF) is a Linux kernel subsystem that enables verified, sandboxed programs to run in kernel space without requiring kernel modification~\cite{rice_learning_2023}. eBPF programs attach to kernel events, such as network packet processing or system calls, and execute directly within the kernel. This approach eliminates the overhead associated with user-kernel context switches that are prevalent in proxy-based architectures. Prior to execution, the kernel verifier guarantees that each program is safe, ensuring completion within a bounded time, and preventing system crashes or instability.

\noindent\textbf{Cilium.} Cilium is a production service mesh and container networking plugin built on eBPF~\cite{noauthor_cilium_nodate}, and has become the dominant CNI in production Kubernetes deployments, used in over 60\% of surveyed clusters and adopted by all major public cloud providers~\cite{cilium_annual_2025,cilium_journey_2024}. Unlike sidecar-based meshes, Cilium enforces Layer~3 and Layer~4 policies entirely within the kernel via eBPF, requiring no per-pod proxy. A single Cilium agent runs on each node, managing networking and policy enforcement for all pods on that node by attaching eBPF programs to the kernel's networking stack. For Layer~7 enforcement, however, eBPF's verifier constraints limit the complexity of programs it can execute~\cite{sidoretti2023application,vieira_fast_2021}, so Cilium falls back to a shared Envoy proxy per node, redirecting traffic through user space with the same parsing overhead described above. This hybrid architecture, eBPF for transport-layer enforcement, Envoy as a fallback for Layer~7, makes Cilium a natural foundation for \systemname{}, as discussed in Section~\ref{sec:design}.

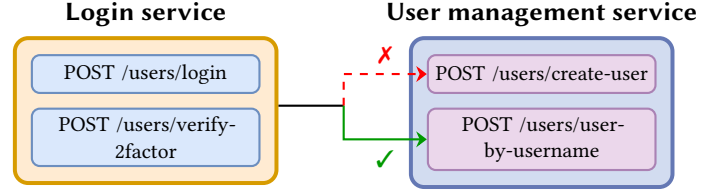
\begin{figure}[tp]
\centering
\begin{tikzpicture}[
  font=\footnotesize,
  node distance=0.18cm and 0.3cm,
  ep/.style={
    rectangle, rounded corners=3pt,
    text width=2.8cm,
    minimum height=0.5cm,
    align=center,
    draw, thick,
    inner sep=3pt,
  },
  arr/.style={-{Stealth[length=4pt,width=6pt,line width=1pt]}, thick},
]

\node[ep, fill=epBlueFill, draw=epBlueBdr]
  (lep1) {{{POST /users/login}}};

\node[ep, fill=epBlueFill, draw=epBlueBdr, below=of lep1]
  (lep2) {{{POST /users/verify-2factor}}};

\begin{pgfonlayer}{background}
  \node[draw=loginBorder, line width=1.5pt,
        rounded corners=5pt, fill=loginFill,
        fit=(lep1)(lep2), inner sep=6pt,
        label={[font=\sffamily\bfseries]
               above:{Login service}}]
    (loginBox) {};
\end{pgfonlayer}

\node[ep, fill=epPinkFill, draw=epPinkBdr,
      right=2.2cm of lep1]
  (rep1) {{{POST /users/create-user}}};

\node[ep, fill=epPinkFill, draw=epPinkBdr, below=of rep1]
  (rep2) {{{POST /users/user-by-username}}};

\begin{pgfonlayer}{background}
  \node[draw=userBorder, line width=1.5pt,
        rounded corners=5pt, fill=userFill,
        fit=(rep1)(rep2), inner sep=6pt,
        label={[font=\sffamily\bfseries]
               above:{User management service}}]
    (userBox) {};
\end{pgfonlayer}

\coordinate (midEp) at ($(lep1)!0.5!(lep2)$);
\coordinate (exitPt) at (loginBox.east |- midEp);
\coordinate (splitPt) at ($(exitPt)!0.5!(userBox.west |- midEp)$);

\draw[thick] (exitPt) -- (splitPt);

\draw[arr, red, dashed]
  (splitPt) -- (splitPt |- rep1.west) -- (rep1.west);

\node[font=\small\bfseries, text=red]
  at ($(splitPt |- rep1.west)!0.5!(rep1.west) + (0,0.25cm)$)
  {\ding{55}};

\draw[arr, green!60!black]
  (splitPt) -- (splitPt |- rep2.west) -- (rep2.west);

\node[font=\small\bfseries, text=green!60!black]
  at ($(splitPt |- rep2.west)!0.5!(rep2.west) + (0,-0.25cm)$)
  {\ding{51}};

\end{tikzpicture}

\caption{%
A microservice interaction scenario illustrating endpoint-level access within a single service-to-service communication channel.
The login service legitimately calls \texttt{POST /users/user-by-username} (\textcolor{green!60!black}{\ding{51}}), but a compromised instance could invoke \texttt{POST /users/create-user} to create a backdoor account (\textcolor{red}{\ding{55}}). Transport-layer controls cannot distinguish these endpoints; Layer~7 enforcement is required.
}
\label{fig:example}
\end{figure}

\subsection{The Need for Fine-Grained Policy Enforcement}
Microservice applications consist of services that communicate over networked APIs, often exposing multiple logically distinct operations through a single service interface~\cite{DeathStarBench-ASPLOS19,managedService-NSDI23}. Consequently, the microservice architectural style increases the attack surface and requires careful control over service-to-service interactions~\cite{SoK-ARES22,istio_security,google_asm_security,cilium_security}. Figure~\ref{fig:example} illustrates a simplified login workflow in such a system.

The login service is responsible for authenticating users. To fulfill this role, it invokes two endpoints \emph{POST /users/login} from the login service and \emph{POST /users/user-by-username}  on the user management service. This microservice call verifies credentials and retrieves the corresponding account record. However, the same service also exposes an additional endpoint, such as \emph{POST /users/create-user}, which are unrelated to the login functionality.

If the login service is compromised, an adversary can exploit its existing network connectivity to invoke these additional endpoints. For example, issuing a request to \emph{/users/create-user} enables the creation of a persistent backdoor account. This represents a form of lateral movement, where a compromised service abuses legitimate communication channels to perform unintended actions~\cite{ncsc_lateral_movement,SoK-ARES22}.

Enforcing the principle of least privilege would prevent such attacks by restricting the login service to only the endpoints required for its operation.  In practice, modern service meshes support such
fine-grained authorization policies~\cite{istio_security,cilium_security}. However, transport-layer (Layer~4) policies operate at the granularity of IP addresses and ports, and thus cannot distinguish between different API operations multiplexed over the same connection~\cite{zhu_dissecting_2023}. In this example, requests to \emph{/users/login}, \emph{/users/user-by-username}, and \emph{/users/create-user} are indistinguishable at Layer~4, as they share the same destination port.

Supporting least-privilege enforcement in microservice environments, therefore, requires application-layer (Layer~7) visibility, where policies can match on request semantics such as HTTP method and path. This motivates the need for fine-grained policy enforcement mechanisms that operate beyond the transport layer.

\begin{lstlisting}[
  float,
  floatplacement=H,
  language=YAML,
  label=lst:cnp-login,
  caption={%
    \texttt{CiliumNetworkPolicy} enforcing least-privilege access from the login service to the user management service. 
    }]
apiVersion: cilium.io/v2
kind: CiliumNetworkPolicy
metadata:
  name: policy-user-mgmt-service
spec:
  endpointSelector:
    matchLabels:
      app: user-management-service
  ingress:
    - fromEndpoints:
        - matchLabels:
            app: login-service
      toPorts:
        - ports:
            - port: "8080"
              protocol: TCP
          rules:
            http:
              - method: POST
                path: /api/v1/users/user-by-username
\end{lstlisting}

Listing~\ref{lst:cnp-login} shows what this policy looks like as a \emph{CiliumNetworkPolicy}. The \emph{endpointSelector} targets the user management service; the \emph{http} rule admits only the path the login service legitimately needs. Any other request from the login service, including \emph{POST /users/create-user}, is denied by the implicit default-deny behavior of the policy.

To evaluate this policy, Cilium must redirect traffic from the login service through a per-node Envoy proxy, which fully parses the HTTP request, matches the path against the rules, and forwards or drops the request before the application receives it. As the number of services and endpoints grows, this proxy-based evaluation applies to every inter-service request on every hop of every call path. 

\begin{figure}[tp]
\centering
\begin{tikzpicture}[
  font=\small,
  node distance=0.45cm and 0.3cm,
  layer/.style={
    rectangle, rounded corners=3pt,
    minimum width=2.4cm, minimum height=0.52cm,
    text centered, draw, thick
  },
  kernelbox/.style={layer,
    fill=blue!12, draw=blue!50!black},
  proxybox/.style={layer,
    fill=orange!20, draw=orange!70!black},
  appbox/.style={layer,
    fill=green!15, draw=green!50!black},
  redbox/.style={layer,
    fill=red!18, draw=red!70!black,
    text=red!80!black},
  arr/.style={-{Stealth[length=4pt]}, thick},
  redarr/.style={arr, red!70!black, dashed},
  seclabel/.style={font=\bfseries\small},
  annot/.style={font=\small, text=gray},
]

\begin{scope}[local bounding box=L4scope]
 
  \node[seclabel] (l4title)
        {\textbf{(a) Layer~4 (eBPF)}};
 
  \node[kernelbox, below=0.5cm  of l4title] (l4net)
        {Read from TCP stream};
  \node[kernelbox, below=0.52cm of l4net]   (l4pol)
        {IP/port policy check (eBPF)};
  \node[appbox,    below=0.52cm of l4pol]   (l4app)
        {Application};
 
  \draw[arr] (l4net.south) -- (l4pol.north);
  \draw[arr] (l4pol.south)
        -- node[right, annot]{allow / deny}
        (l4app.north);
 
  \node[annot, below=0.2cm of l4app, align=center]
        {\textbf{3 steps} --- no parsing};
 
\end{scope}

\begin{scope}[local bounding box=L7scope,
              xshift=5.2cm]

  \node[seclabel] (l7title)
        {\textbf{(b) Layer~7 (proxy)}};

  \node[kernelbox, below=0.5cm  of l7title]      (l7net)
        {Read from TCP stream};
  \node[redbox,    below=0.52cm of l7net]         (l7unmarshal)
        {Parse request (proxy)};
  \node[proxybox,  below=0.52cm of l7unmarshal]   (l7inspect)
        {Inspect + policy check};
  \node[redbox,    below=0.52cm of l7inspect]     (l7marshal)
        {Serialize to TCP stream};
  \node[redbox,    below=0.52cm of l7marshal]     (l7unmarshal2)
        {Parse request (app)};
  \node[appbox,    below=0.52cm of l7unmarshal2]  (l7app)
        {Application logic};

  \draw[arr]    (l7net.south)        -- (l7unmarshal.north);
  \draw[arr]    (l7unmarshal.south)  -- (l7inspect.north);
  \draw[arr]    (l7inspect.south)    -- (l7marshal.north);
  \draw[redarr] (l7marshal.south)    -- (l7unmarshal2.north);
  \draw[arr]    (l7unmarshal2.south) -- (l7app.north);

  \draw[decorate,
        decoration={brace, amplitude=6pt, mirror},
        line width=1pt, orange!70!black]
    ($({l7unmarshal.west} -| {l7marshal.west}) + (-0.1cm, 0.26cm)$) --
    ($(l7marshal.west) + (-0.1cm, -0.26cm)$)
    node[midway, left=7pt, annot,
         text=orange!80!black, align=right]
         {Envoy\\proxy};

  \node[annot, left=0.12cm of l7unmarshal2,
        text=red!70!black, align=right]
        {redundant};

  \node[annot, below=0.2cm of l7app, align=center]
        {\textbf{6 steps} --- 2$\times$ parse, \\IPC overhead};

\end{scope}

\draw[dashed, gray!50, thick]
  (2.2, 0.9) -- (2.2, -8.6);

\node[layer, fill=gray!12, draw=gray!60,
      above=0.5cm of l4title,
      xshift=2.6cm]
      (sender) {Sender service};

\draw[arr] (sender.south west)
      to[out=230, in=110]
      ([xshift=0.4cm]l4title.north);
\draw[arr] (sender.south east)
      to[out=310, in=70]
      ([xshift=-0.4cm]l7title.north);

\end{tikzpicture}
\caption{%
  Comparison of Layer~4 and Layer~7 policy enforcement for a
  single inter-service request.
  \textbf{(a)}~Layer~4 enforcement via eBPF and
  \textbf{(b)}~Layer~7 enforcement via a sidecar proxy.
}
\label{fig:l4vsl7}
\end{figure}
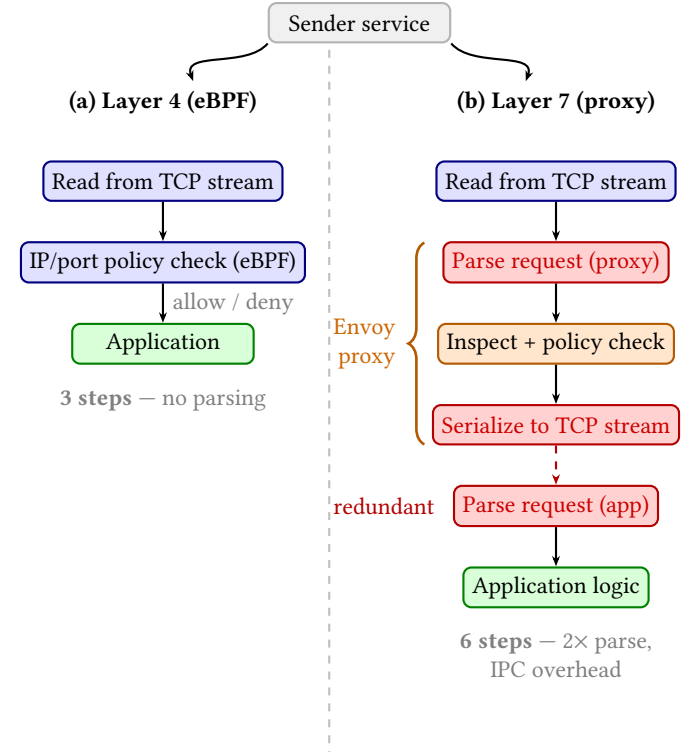

\subsection{Performance Implications of Layer~7 Inspection}

Although the necessity for fine-grained Layer~7 policy enforcement is well established, achieving this efficiently in practice remains a challenge. Existing service meshes implement such policies by intercepting all inter-service traffic through a dedicated proxy, which parses each request at the application layer to evaluate applicable policies before forwarding. Figure~\ref{fig:l4vsl7} illustrates the contrast with transport-layer enforcement: a Layer~4 eBPF check requires no parsing and forwards the request directly to the application, whereas a Layer~7 proxy must unmarshal, inspect, re-marshal, and forward the request, resulting in the application parsing the same request a second time. This interception model is architecture-agnostic and does not require modifications to individual services, which supports broad deployment. However, it introduces per-request overhead that accumulates along every hop of a multi-service request path, which are typically the most latency-sensitive in production deployments~\cite{dean_tail_2013,HybridMesh-NSDI26}. 

To quantify this overhead, we measure the end-to-end latency of Cilium under two configurations: (i) transport-layer enforcement implemented entirely in the eBPF data path, and (ii) Layer~7 enforcement achieved by redirecting traffic through the Envoy proxy.

\begin{table}[tp]
  \centering
  \caption{%
    End-to-end latency overhead of Cilium under the transport-layer and Layer~7 enforcement modes, measured in our evaluation (Section~\ref{section:results_e2e_latency}).
  }
  \label{tab:sidecar_overhead}
  \resizebox{\linewidth}{!}{
  \begin{tabular}{lcccc}
    \toprule
    \textbf{Mode}
      & \textbf{Latency} & \textbf{CPU}
      & \textbf{Parsing share} & \textbf{Parsing share} \\
      & \textbf{overhead} & \textbf{overhead}
      & \textbf{(latency)} & \textbf{(CPU)} \\
    \midrule
    TCP proxy   & 1$\times$ & 1$\times$
                & $<$10\%           & $<$10\%         \\
    HTTP proxy  & 3$\times$ & 3$\times$
                & 62--77\%          & $>$60\%         \\
    \bottomrule
  \end{tabular}}
\end{table}
Table~\ref{tab:sidecar_overhead} shows that transport-layer-only enforcement incurs negligible overhead. Enabling Layer~7 inspection, however, increases end-to-end latency by approximately 3$\times$ and CPU consumption by a comparable factor. A detailed breakdown attributes 62--77\% of the latency increase and more than 60\% of the additional CPU utilization to protocol parsing alone.

These results are consistent with prior analyses of service mesh data planes. In particular, Zhu et al.~\cite{zhu_dissecting_2023} identify proxy-based protocol parsing as the dominant contributor to both latency and CPU overhead under Layer~7 enforcement. The consistency between our measurements and prior work strengthens the conclusion that protocol parsing is the primary source of service-mesh overhead and therefore the key target for optimization.

\noindent\textbf{Adoptability.} The performance overhead of proxy-based enforcement has prompted several efforts to move policy decisions closer to the application. At the library level, ServiceRouter~\cite{ServiceRouter-OSDI23} embeds service mesh functionality directly into each service, eliminating the need for a proxy and achieving low latency and resource consumption. At the kernel level, recent work explores offloading Layer~7 processing into eBPF~\cite{sidoretti2023application} or kernel modules to remove the user-space proxy from the data path altogether. 

These approaches fundamentally trade deployability for performance. Library-based designs require applications to be modified and recompiled against a shared runtime, while kernel-level solutions depend on custom kernel extensions deployed on every node. In practice, production environments are heterogeneous, often including legacy services, third-party containers, and workloads with limited developer control. In such settings, imposing invasive changes to either application code or the kernel is rarely feasible. As a result, operators default to proxy-based service meshes, accepting their performance overhead in exchange for deployability.

This gap motivates an architecture that achieves application-level performance without requiring universal adoption or infrastructure changes. Cilium provides a suitable foundation: its eBPF data plane handles transport-layer enforcement, and its existing Layer~7 Envoy proxy serves as a fallback for services that cannot be modified. \systemname{} builds on this foundation, making application-level enforcement \emph{opt-in} rather than mandatory, so that services adopt it incrementally without disrupting the rest of the deployment.

\section{Design}\label{sec:design}
We design \systemname to bridge the gap between tightly coupled, high-performance service frameworks and broadly deployable but proxy-heavy service meshes. \systemname targets deployments that require service-mesh security guarantees while avoiding the latency overhead of per-pod Layer~7 proxies. By shifting Layer-7 policy enforcement into an optional application-bound library, \systemname eliminates redundant parsing and removes the proxy from the critical data path whenever possible. At the same time, by integrating with Cilium, \systemname maintains compatibility with existing Kubernetes deployments and supports a graceful fallback to proxy-based enforcement for services that cannot be modified.

In doing so, \systemname provides a practical, deployable architecture that significantly reduces latency overhead while maintaining the security guarantees and operational assumptions expected from a service mesh. This combination of performance, deployability, and compatibility with existing service-mesh assumptions motivates a new design point that departs from traditional sidecar-based architectures.

\subsection{Architecture}

\begin{figure}[tp]
    \centering
    \includegraphics[width=0.89\linewidth]{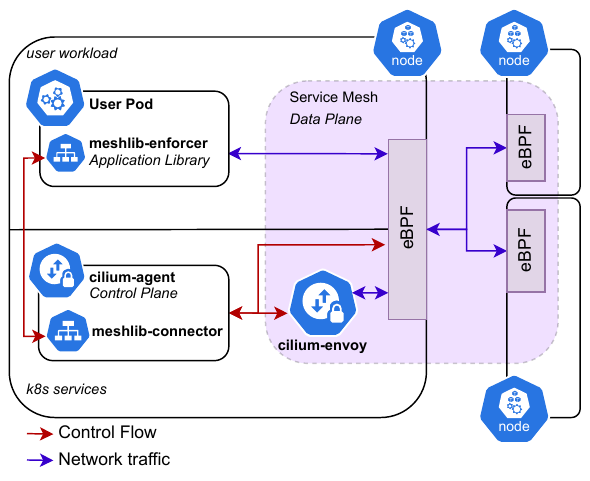}
    \caption{The architecture of \systemname, integrated within a user-specified pod using the \systemname{}-enforcer library, and the \systemname{}-connector within Cilium's control plane}
    \label{fig:meshlib-arch}
\end{figure}

Figure~\ref{fig:meshlib-arch} illustrates the architecture of \systemname and highlights the minimal changes required to remove Layer~7 proxies from the critical data path. Rather than replacing the service mesh, \systemname introduces an application-bound enforcement component while reusing Cilium’s existing control and data planes.

\systemname is composed of two main components: (i) an application-bound library that performs policy enforcement at the application layer, and (ii) an extension to Cilium’s control plane that coordinates configuration and integration with the existing Cilium data plane. Together, these components extend Cilium’s service mesh architecture without modifying its core data-path behavior.

\subsubsection{Application Library}

The application library (\emph{\systemname{}-enforcer}) is embedded directly into microservices that opt in to low-latency, application-layer policy enforcement. It operates at the application layer (e.g., HTTP) and reuses the protocol parsing that is already performed by the application framework.
By doing so, it avoids the redundant parsing and inter-process communication overhead introduced by traditional sidecar proxies.

The library integrates with the receiving side of the communication, where it is attached to the HTTP server running inside the application container. All incoming requests are evaluated against the applicable security policies before being passed to the application logic. This ensures that only authorized and trusted traffic reaches the service while preserving the same trust assumptions and policy semantics as a traditional service mesh.

To obtain the policies it enforces, the application library communicates with the \systemname connector in Cilium's control plane over a trusted channel. The received configuration includes the relevant application-layer policies and identity information required to evaluate incoming requests. Once initialized, the library enforces these policies locally within the application process.

The application library is optional and non-intrusive. Services that do not include the \systemname library remain unchanged and continue to rely on Cilium's existing service mesh behavior.

\subsubsection{Control Plane Extension}

\systemname-connector is implemented as an extension to Cilium's control plane. It retains authority over application-layer policy distribution and coordinates enforcement decisions with \systemname-enabled services.

This component integrates with Cilium's existing policy and identity infrastructure. It determines which services are \systemname-enabled and provides the corresponding policy configuration to their application libraries. For these services, the connector instructs Cilium to bypass its Layer 7 proxy for the relevant traffic flows, allowing requests to be delivered directly to the application. In contrast, Layer 3 and Layer 4 enforcement remains handled by Cilium's eBPF-based data plane.

For services that do not embed \systemname{}'s application library, Cilium continues to enforce policies using its standard mechanisms, including the Layer 7 proxy when required. This hybrid operation preserves backward compatibility while allowing \systemname{} to be deployed incrementally, enabling sidecar-less enforcement only for services that explicitly opt in.

\subsection{The \systemname{} Protocol}\label{subsec:protocol}
The \systemname{} protocol coordinates policy enforcement state between an application-bound library and the local Cilium agent. Its design enforces the following invariant:
\emph{Traffic is admitted to in-process enforcement only after the
library has fully installed and acknowledged the current policy
version; otherwise, enforcement remains on the proxy path.}

To enforce this invariant, the protocol provides two tightly coupled functions: \emph{policy distribution}, which delivers authoritative Layer-7 access control state, and \emph{identity resolution}, which maintains a consistent mapping between pod IP addresses and Cilium security identities. Both functions are multiplexed over a single persistent bidirectional gRPC stream, allowing policy updates and identity changes to be applied asynchronously without per-request control-plane overhead.

\subsubsection{Service Interface}
Listing~\ref{lst:proto-service} defines the service.
The library opens a single \emph{PolicyStream} connection to the local Cilium agent on startup. 
The \texttt{stream} qualifier on \emph{both} the request and response types declares a fully bidirectional channel: unlike a unary RPC, either side may send messages at any time without waiting for a reply. This allows the agent to push policy updates proactively and allows the library to send acknowledgments independently, with no coupling between the two directions. The stream remains open for the pod's lifetime; all messages are multiplexed over it using \texttt{oneof} discriminators, so new message types can be added without breaking backwards compatibility.

\begin{lstlisting}[language=protobuf,
  label=lst:proto-service,
  caption={The \texttt{PolicyEnforcer} gRPC service.
    One persistent bidirectional stream carries all
    library$\to$agent and agent$\to$library messages.}]
service PolicyEnforcer {
  rpc PolicyStream(stream PolicyStreamRequest)
      returns (stream PolicyStreamResponse);
}
\end{lstlisting}

\subsubsection{Library-to-Agent Messages} 
Library-to-agent messages signal enforcement readiness and confirm policy installation. They allow the Cilium agent to precisely control when traffic transitions away from the proxy path, preventing enforcement gaps during startup, updates, and teardown. 

The library emits three message types (Listing~\ref{lst:proto-request}). Each message includes a \texttt{port} field identifying the application's inbound TCP port (e.g., 8080 or 12340), distinct from the gRPC control channel. A single library instance may manage multiple ports by issuing separate messages per port: \circnum{1}~\textbf{EnforceReady}, which signals readiness to enforce policies on a port and allows Cilium to divert traffic from the Envoy proxy; \circnum{2}~\textbf{EnforceTerminate}, which relinquishes enforcement so that Cilium restores the proxy path during graceful shutdown; and \circnum{3}~\textbf{EnforceAck}, which acknowledges a specific policy version, confirming that it has been installed before traffic is admitted.

\begin{lstlisting}[float,
  floatplacement=H,language=protobuf,
  label=lst:proto-request,
  caption={Library-to-agent request messages. Exactly one
    variant is set per frame.}]
message PolicyStreamRequest {
  oneof request_type {
    InboundEnforceReadyMessage     inbound_enforce_ready     = 1;
    InboundEnforceTerminateMessage inbound_enforce_terminate = 2;
    InboundEnforceAckMessage       inbound_enforce_ack       = 3;
  }
}
message InboundEnforceReadyMessage     { uint32 port = 1; }
message InboundEnforceTerminateMessage { uint32 port = 1; }
message InboundEnforceAckMessage  { uint32 port     = 1;
                                    uint32 sequence = 2; }
\end{lstlisting}

\subsubsection{Agent-to-Library Messages} 
Agent-to-library messages carry the authoritative enforcement state. The agent emits exactly two message types. The \emph{InboundEnforceStateMessage} conveys the policy rules, while the \emph{IPCacheUpdateMessage} provides the IP-to-identity mapping required to determine which policy applies to each incoming connection. 

Decoupling policy state from identity updates avoids unnecessary policy re-evaluation: identity mappings change frequently as pods are created and removed, whereas policy rules are relatively stable. Listing~\ref{lst:proto-response} shows the response envelope.

\noindent\textbf{Policy state (\emph{InboundEnforceStateMessage}).} This message carries the authoritative policy state for an inbound port and is the only mechanism through which the agent installs or updates policies in the library.

\begin{lstlisting}[float,
  floatplacement=H,language=protobuf,
  label=lst:proto-response,
  caption={Agent-to-library response messages.}]
message PolicyStreamResponse {
  oneof response_type {
    InboundEnforceStateMessage inbound_enforce_state = 1;
    IPCacheUpdateMessage       ipcache_update        = 2;
  }
}
\end{lstlisting}

The fields of this message are as follows. \texttt{port} identifies the inbound port for which this policy state applies, matching the port registered by \texttt{EnforceReady}. \texttt{sequence} is a monotonically increasing version counter echoed in \texttt{EnforceAck}. \texttt{should\_enforce = false} passes all traffic without inspection during policy transitions, without touching eBPF maps. Each \texttt{Policy} is scoped to Cilium numeric \emph{security identities} (derived from pod labels, stable across restarts) and a list of \texttt{HTTPRule} matchers on \texttt{path} (regex), \texttt{method}, \texttt{host}, and \texttt{headers}; absent fields match any value.

Each \texttt{Policy} entry specifies a \textbf{\texttt{type}}, which can be either \texttt{ALLOW} or \texttt{DENY}. An \texttt{ALLOW} entry authorizes matching requests to proceed to the application, while a \texttt{DENY} entry explicitly rejects matching requests and returns \texttt{HTTP~403} prior to the invocation of application code. Explicit \texttt{DENY} entries are employed to define exceptions within broader \texttt{ALLOW} rules. The \textbf{\texttt{endpoint\_identities}} field contains a list of Cilium numeric security identities, where each identity is a \texttt{uint32} assigned by Cilium to a group of pods with identical Kubernetes labels. These identities remain consistent across pod restarts; although a pod's IP address may change upon rescheduling, its identity remains unchanged. The library determines the identity of each incoming connection's source IP by consulting the local IP cache before selecting the appropriate \texttt{Policy} entry.

\noindent\textbf{\texttt{http\_rules}} specifies a set of \texttt{HTTPRule} match conditions. A rule is satisfied when all specified fields match the request, while unspecified fields act as wildcards.

Each \texttt{HTTPRule} can constrain multiple request attributes: \texttt{path} is a regular expression over the URI path, \texttt{method} denotes the HTTP verb (e.g., \texttt{GET}, \texttt{POST}), and \texttt{host} matches the \texttt{Host} header, which is useful when a service exposes multiple virtual hosts on the same port. The \texttt{headers} field defines a set of required headers in \texttt{"Name: Value"} format; all listed headers must be present for the rule to match.

A \texttt{Policy} matches a request if the source identity appears in \texttt{endpoint\_identities} and at least one \texttt{HTTPRule} in \texttt{http\_rules} is satisfied. Thus, matching follows conjunction (AND) across fields within a rule and disjunction (OR) across rules within a policy. Listing~\ref{lst:proto-state} shows the complete message definition.

\begin{lstlisting}[float,
  floatplacement=H,language=protobuf,
  label=lst:proto-state,
  caption={Policy state message. \texttt{should\_enforce=false}
    signals that Cilium has reclaimed enforcement.}]
message InboundEnforceStateMessage {
  uint32          port           = 1;
  uint32          sequence       = 2;
  bool            should_enforce = 3;
  repeated Policy policies       = 4;

  message Policy {
    enum Type { DENY = 0; ALLOW = 1; }
    Type              type                = 1;
    repeated uint32   endpoint_identities = 2;
    repeated HTTPRule http_rules          = 3;

    message HTTPRule {
      optional string path    = 1;
      optional string method  = 2;
      optional string host    = 3;
      repeated string headers = 4;
    }
  }
}
\end{lstlisting}

\noindent\textbf{Identity resolution (\texttt{IPCacheUpdateMessage}).} Policies are expressed in terms of Cilium security identities rather than IP addresses. As a result, the library must resolve the source IP of each incoming connection to its corresponding identity before evaluating policies. This mapping is maintained by the agent, which streams \texttt{IPCacheUpdateMessage} updates to the library.

The update mechanism supports both full and incremental synchronization. When \texttt{should\_clear} is set, the library discards its local cache and replaces it with the entries provided in \texttt{upsert\_entries}. This occurs during initialization or after disruptive events such as agent restarts or node reconnections. During steady-state operation, \texttt{should\_clear} is unset, and updates consist only of incremental changes conveyed through \texttt{upsert\_entries} and \texttt{delete\_entries}.

\texttt{upsert\_entries} introduce or update mappings from IP addresses to identities. Each \emph{IPUpsertEntry} encodes the IP address in raw binary form (4 bytes for IPv4, 16 bytes for IPv6), avoiding string parsing overhead in this high-frequency path. The \texttt{ip\_type} field distinguishes between \texttt{IPV4} and \texttt{IPV6}, enabling support for dual-stack deployments. \texttt{delete\_entries} remove stale mappings when pods terminate or are rescheduled.

At enforcement time, the library performs a lookup on the source IP of each accepted connection. If a mapping is found, the corresponding identity is used for policy evaluation. If no mapping exists, for example, during the brief window before an update is received, the connection is treated as originating from an unknown identity. Such requests match no \texttt{ALLOW} policy and are therefore denied by default.

\begin{lstlisting}[float,
  floatplacement=H,language=protobuf,
  label=lst:proto-ipcache,
  caption={\texttt{IPCacheUpdateMessage}. \texttt{should\_clear=true}
    triggers a full table replacement on initialisation.}]
message IPCacheUpdateMessage {
  bool                   should_clear   = 1;
  repeated IPUpsertEntry upsert_entries = 2;
  repeated IPDeleteEntry delete_entries = 3;
}
message IPUpsertEntry { IPType ip_type=1; bytes ip=2;
                        uint32 identity=3; }
message IPDeleteEntry { IPType ip_type=1; bytes ip=2; }
enum IPType { IPV4 = 0; IPV6 = 1; }
\end{lstlisting}

\subsubsection{Protocol Lifecycle}

\begin{figure}[t]
\centering

\begin{tikzpicture}[
  x=1cm, y=-1cm,
  font=\fontsize{6.5}{8}\selectfont,
  arr/.style={
    -{Stealth[length=4pt, width=6pt]},
    line width=1pt},
  ll/.style={draw=gray!95, line width=0.8pt, dashed},
  ml/.style={
    font=\fontsize{7}{8.0}\selectfont\ttfamily,
    fill=white, inner sep=0.8pt, midway, above=5.0pt},
  sl/.style={
    font=\fontsize{6.5}{8}\selectfont\itshape,
    text=purple, inner sep=0pt, above=1.0pt},
  sep/.style={
    font=\fontsize{7}{8.5}\selectfont\itshape,
    text=blue},
  pb/.style={
    draw=gray!60, rounded corners=2pt, fill=white,
    inner sep=3pt,line width=0.8pt,
    font=\fontsize{7}{8.5}\selectfont\sffamily\bfseries},
]

\def\LX{0.0}   
\def\RX{5.6}   
\def\MX{2.8}   

\node[pb, fill=orange!15, draw=orange!50!black,font=\small] (lib)    at (\LX, 0) {Library};
\node[pb, fill=orange!15, draw=orange!50!black,font=\small] (cilium) at (\RX, 0) {Cilium agent};

\draw[ll] (\LX, 0.38) -- (\LX, 9.55);
\draw[ll] (\RX, 0.38) -- (\RX, 9.55);

\draw[arr] (\LX,1.0) -- (\RX,1.0)
  node[ml]{PolicyStream()};
\node[sl, anchor=north west] at (\LX, 0.82) {(1) open stream};

\draw[arr] (\LX,2.0) -- (\RX,2.0)
  node[ml]{EnforceReady(port=8080)};
\node[sl, anchor=north west] at (\LX, 1.82) {(2) register port};

\draw[arr] (\RX,3.0) -- (\LX,3.0)
  node[ml]{IPCacheUpdate(clear=true, \ldots)};
\node[sl, anchor=north east] at (\RX, 2.82) {(3) seed IP cache};

\draw[arr] (\RX,4.0) -- (\LX,4.0)
  node[ml]{EnforceState(seq=1, enforce=true, policies)};
\node[sl, anchor=north east] at (\RX, 3.82) {(4) push policies};

\draw[arr] (\LX,5.0) -- (\RX,5.0)
  node[ml]{EnforceAck(seq=1)};
\node[sl, anchor=north west] at (\LX, 4.82) {(5) acknowledge};

\node[sep] at (\MX, 5.55)
  {\textbf{$\leftarrow$ steady state: library enforces in-process $\rightarrow$}};

\draw[arr] (\RX,6.2) -- (\LX,6.2)
  node[ml]{EnforceState(seq=2, enforce=true, policies')};
\node[sl, anchor=north east] at (\RX, 6.02) {(6) policy update};

\draw[arr] (\LX,7.1) -- (\RX,7.1)
  node[ml]{EnforceAck(seq=2)};
\node[sl, anchor=north west] at (\LX, 6.92) {(7) acknowledge};

\draw[arr] (\RX,8.0) -- (\LX,8.0)
  node[ml]{IPCacheUpdate(upsert: 10.0.1.5$\to$id=42)};
\node[sl, anchor=north east] at (\RX, 7.82) {(8) new pod};

\draw[arr] (\LX,9.0) -- (\RX,9.0)
  node[ml]{EnforceTerminate(port=8080)};
\node[sl, anchor=north west] at (\LX, 8.82) {(9) teardown};

\end{tikzpicture}
\caption{%
  Protocol lifecycle for one inbound port.
  The library opens a persistent stream~(1), registers the
  port~(2), and receives the initial IP cache~(3) and policy
  set~(4).
  It acknowledges to activate enforcement~(5).
  Policy changes~(6--7) and IP cache deltas~(8) arrive on the
  same stream.
  On teardown~(9), Cilium reinstates the Envoy proxy for
  that port, ensuring continuous enforcement.
}
\label{fig:proto-lifecycle}
\end{figure}
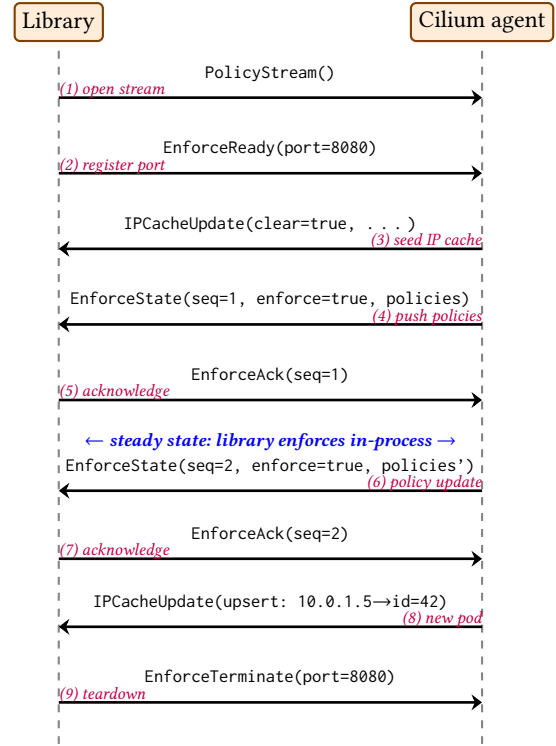

Figure~\ref{fig:proto-lifecycle} illustrates the protocol lifecycle for a single inbound port. Enforcement is enabled only after the library acknowledges receipt of the initial policy state. Subsequent policy updates and identity changes are delivered asynchronously on the same stream, while acknowledgements ensure that each update is fully installed before it becomes effective.

\subsubsection{Concrete Example}

Listing~\ref{lst:proto-example} shows a concrete \emph{InboundEnforceStateMessage} for \emph{ts-auth-service} from TrainTicket benchmark~\cite{zhou_benchmarking_2018}. The agent enforces two \emph{ALLOW} policies on port~12340: the gateway service (identity~1234) may call five endpoints; the user service (identity~5678) may call only the authentication endpoint. All other requests are implicitly denied.

\begin{lstlisting}[float,
  floatplacement=H,language=protobuf,
  label=lst:proto-example,
  caption={Concrete \texttt{InboundEnforceStateMessage} for
    \texttt{ts-auth-service} (port~12340, seq~1).
    Two \texttt{ALLOW} policies are installed; requests from
    any identity not listed in either policy are denied by
    default.}]
InboundEnforceStateMessage {
  port=12340  sequence=1  should_enforce=true
  policies=[
    Policy {                        // ts-gateway-service
      type=ALLOW  endpoint_identities=[1234]
      http_rules=[
        HTTPRule{method="GET",    path="/api/v1/users/hello.*"},
        HTTPRule{method="POST",   path="/api/v1/users/login.*"},
        HTTPRule{method="DELETE", path="/api/v1/users/.*"},
        HTTPRule{method="GET",    path="/api/v1/auth/hello.*"},
        HTTPRule{method="POST",   path="/api/v1/auth/.*"},
      ]
    },
    Policy {                        // ts-user-service
      type=ALLOW  endpoint_identities=[5678]
      http_rules=[
        HTTPRule{method="POST", path="/api/v1/auth/.*"},
      ]
    }
  ]
}
\end{lstlisting}

On receipt, the library sends \texttt{EnforceAck(seq=1)}. Each subsequent request on port~12340 is evaluated in-process: the library resolves the source IP to a Cilium identity via the local cache, selects the matching \texttt{Policy}, and checks the \texttt{http\_rules} against the already-parsed HTTP request object. Matching \texttt{ALLOW} rules proceed to the application; all others return \texttt{HTTP~403} before application code is invoked.

\subsubsection{Design Rationale}
The protocol design reflects four goals: minimizing control-plane overhead, preserving policy correctness during updates, supporting incremental deployment, and aligning with Cilium’s identity model.

\noindent\textbf{\circnum{1}~Single persistent stream.} Each pod establishes one long-lived gRPC stream that is opened once and reused throughout its lifetime. This design eliminates per-request connection setup overhead and enables Cilium to deliver policy updates asynchronously, avoiding the need for polling by the library.

\noindent\textbf{\circnum{2}~Identity-based policy scoping.} Policies are expressed in terms of Cilium security identities rather than IP addresses. Because these identities are derived from stable pod labels, they remain unchanged across pod restarts and rescheduling events, eliminating the need to regenerate policies whenever pod IPs change. At runtime, the \texttt{IPCacheUpdateMessage} dynamically resolves the mapping between stable identities and ephemeral IP addresses.

\noindent\textbf{\circnum{3}~Sequence numbers and acknowledgements.}
Sequence numbers, combined with \texttt{EnforceAck}, provide a lightweight mechanism for confirmed policy delivery and are the basis for enforcing the protocol's no-gap enforcement invariant. Cilium activates a newly installed policy only after receiving the corresponding acknowledgement from the library, ensuring that traffic is never processed under an outdated policy version and preventing race conditions during policy updates.

\noindent\textbf{\circnum{4}~\emph{should\_enforce} flag.}
The \texttt{should\_enforce} flag provides a clean handoff signal: when set to \texttt{false}, it informs the library that Cilium has reclaimed enforcement responsibility because the library is shutting down, the agent connection is being torn down, or the applicable policies no longer require application-level enforcement.

\subsection{Proof of Concept}

We implement a prototype of \systemname{} to validate the feasibility of sidecar-less, application-bound Layer~7 policy enforcement and to evaluate its practical performance implications. Rather than targeting production completeness, the prototype is designed to exercise the core architectural decisions of \systemname{} and to demonstrate that in-process enforcement can be integrated with an existing service mesh without weakening security guarantees.

The prototype consists of two components: (i) an extension to the Cilium control plane, and (ii) an application-bound library embedded within selected microservices. Together, these components implement a hybrid enforcement model in which Cilium retains responsibility for lower-layer networking and identity management, while application services perform request-level authorization locally.

\subsubsection{Control Plane Extension}

The \systemname{} control plane extension is implemented as an add-on to the Cilium agent running on each Kubernetes node. Its primary role is to retain enforcement authority while coordinating policy distribution and state transitions with \systemname-enabled applications.

The extension exposes a local control endpoint that application libraries use to receive policy and identity updates. Communication is implemented using gRPC, which provides a language-neutral interface and supports efficient, long-lived bidirectional streams--an important property given the need to asynchronously deliver policy updates to heterogeneous applications.

The control plane uses a push-based model in which policy and identity updates are delivered asynchronously to \systemname-enabled services. This design reuses Cilium’s existing policy and identity computation, allowing policy semantics and enforcement authority to remain unchanged while shifting only the enforcement location.

\subsubsection{Application Library}

The application library is embedded directly within microservices and is responsible for enforcing Layer~7 authorization decisions in process. For the prototype, we implement the library in Java and integrate it with Spring Boot, which is also used by the benchmark applications.

Within Spring Boot, the library is implemented as an HTTP request filter that intercepts incoming requests before application logic is invoked. This placement allows the library to reuse protocol parsing performed by the framework, avoiding redundant parsing and data copies.
For each incoming request, the library evaluates the request against the current policy using request metadata and the authenticated source identity. Unauthorized requests are rejected with HTTP~403 before application code executes, while authorized requests proceed unchanged.

From an integration perspective, the interaction between the application library and Cilium is minimized. Cilium continues to handle service discovery, routing, and transport-layer enforcement, while the application library performs only Layer~7 authorization using already-parsed request context. This separation ensures that \systemname does not interfere with Cilium's core networking responsibilities, allows services without the library to coexist seamlessly with \systemname-enabled services and easily implemented with other languages or frameworks.

Maintaining a control channel between each \systemname-enabled application and the Cilium agent introduces additional control-plane complexity and creates a dependency on policy availability. This dependency, however, is explicitly non-blocking. If the control channel cannot be established or is temporarily unavailable, Cilium continues to enforce policies via its existing Layer~7 proxy path. Once the channel is established, enforcement can transition to the application library at runtime, after which Cilium reconfigures the data plane to bypass the proxy.

\section{Performance Evaluation}

We evaluate \systemname{} to quantify the cost of in-process Layer~7 enforcement and to compare its performance against representative service mesh architectures. Our evaluation focuses on three questions:

\begin{enumerate}
    \item How does \systemname{} affect end-to-end request latency?
    \item What overhead does \systemname{} introduce relative to
    proxy-based service meshes?
    \item How does eliminating the Layer~7 proxy affect CPU and
    memory consumption?
\end{enumerate}

\noindent\textbf{Experimental setup.} All experiments were conducted on a dedicated Kubernetes node equipped with an AMD Ryzen~7~7700 processor (8 cores), 64\,GB RAM, and K3S~\cite{telenyk_comparison_2021}, a lightweight Kubernetes distribution fully compatible with upstream Kubernetes.

To isolate the overhead introduced by service-mesh enforcement, all application components were deployed on a single node. This configuration eliminates inter-node network variability, allowing observed performance differences to be attributed directly to the service mesh data plane and policy-enforcement mechanisms rather than to physical network effects. The workload, hardware configuration, and deployment topology remain identical across all evaluated systems.

\noindent\textbf{Benchmark application.}
We use the TrainTicket benchmark~\cite{zhou_benchmarking_2018}, a widely used microservice application representative of cloud-native deployments. TrainTicket consists of dozens of interacting services and multiple backing databases connected through HTTP-based APIs, creating a environment for evaluating service-mesh behavior under fine-grained inter-service communication. The application has been adopted extensively in prior systems research as a representative benchmark for microservice architectures and cloud-native runtime evaluation.

Most TrainTicket services are implemented in Java using Spring Boot, which enables direct integration with the \systemname{} library while preserving the application's original communication patterns and deployment structure.

\noindent\textbf{Service mesh configurations.}
In total, four different service mesh configurations will be tested: Cilium with \systemname{}, Cilium unmodified, Istio, and Linkerd. For all four configurations, the same set of security policies will be enforced, with equivalent versions created across the various configuration structures used by the different service meshes. All service meshes are deployed to the K3S node with standard options.

\begin{figure*}[t]
    \centering
    \begin{subfigure}[t]{0.32\textwidth}
        \centering
        \includegraphics[width=\linewidth]{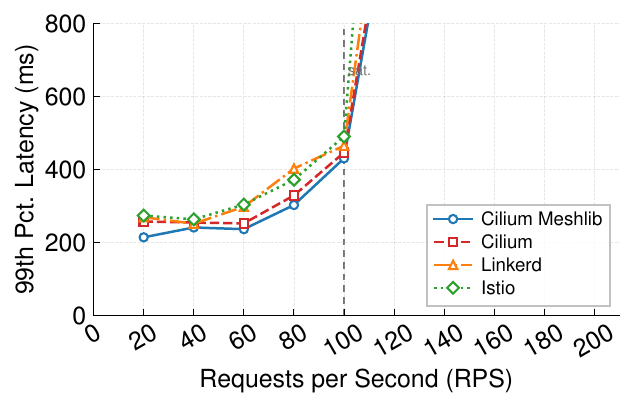}
        \caption{99\%-percentile (ms)}
        \label{fig:result-latency-99p}
    \end{subfigure}
    \begin{subfigure}[t]{0.32\textwidth}
        \centering
        \includegraphics[width=\linewidth]{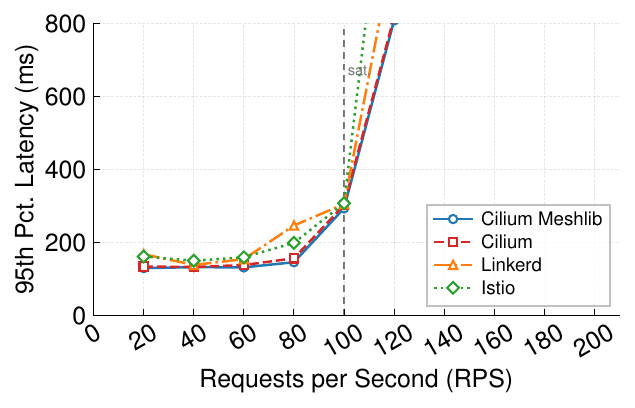}
        \caption{95\%-percentile (ms)}
        \label{fig:result-latency-95p}
    \end{subfigure}
    \begin{subfigure}[t]{0.32\textwidth}
        \centering
        \includegraphics[width=\linewidth]{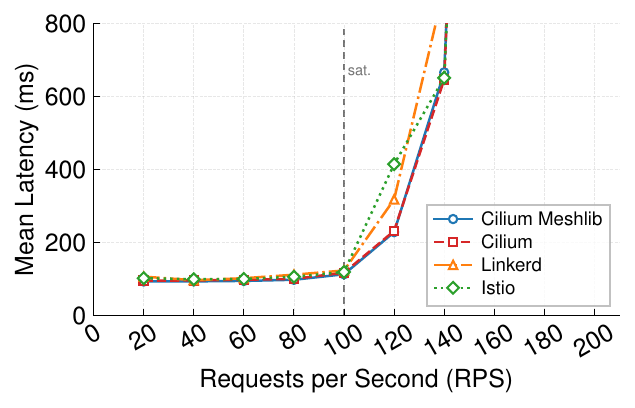}
        \caption{Mean (ms)}
        \label{fig:result-latency-mean}
    \end{subfigure}
    \caption{End-to-end latency of requests to benchmark application for the tested service meshes. The vertical dashed line marks the saturation point at 100~RPS.}
    \label{fig:result-latency}
\end{figure*}

\noindent\textbf{Workload.}
We use wrk2~\cite{tene_wrk2_2019} to generate a closed-loop workload against the TrainTicket application. The workload consists of representative user-facing operations, including login, trip availability queries, and ticket reservation requests. These operations exercise multiple microservices and trigger non-trivial inter-service communication paths, making them suitable for evaluating service-mesh overhead under realistic application behavior.

Unless otherwise stated, wrk2 is configured with 4 worker threads and 16 concurrent connections. We progressively vary the offered load from 20 to 200 requests per second (RPS) to evaluate both steady-state and near-saturation behavior, following common methodology in microservice and datacenter performance evaluations~\cite{DeathStarBench-ASPLOS19}.

\medskip
\noindent\textbf{Policies.}
To ensure a fair comparison, all evaluated service mesh configurations enforce semantically equivalent security policies. Although each mesh uses a different policy language and runtime representation, all configurations implement the same endpoint-level access-control semantics.

Policies were derived from the TrainTicket application's observed service dependencies. For each service-to-service interaction, we generated a least-privilege policy that permits access only to the HTTP endpoints required for correct operation while denying all other requests. This methodology reflects realistic deployment practices in production service meshes and directly exercises Layer~7 enforcement functionality.

In total, the evaluation enforces 126 distinct policies across 37 services, corresponding to an average of 3.4 policies per service. Each policy governs a specific source--destination service pair and restricts access to a subset of the destination service's HTTP endpoints.

\noindent\textbf{Metrics.}
We evaluate the performance of \systemname in three dimensions that directly reflect the cost and effectiveness of its execution policy: (i) end-to-end request latency, which captures the cumulative effect of enforcement along the critical path of microservices; (ii) request overhead, which isolates the communication and policy enforcement costs introduced by the execution model \systemname; and (iii) resource consumption, which quantifies the steady-state cost of enforcing the policy at scale.

\subsection{End-to-end latency}\label{section:results_e2e_latency}

End-to-end latency directly determines user-perceived performance in microservice applications. It captures not only the processing time of the entry-point service, but also the cumulative latency of all inter-service communication along the request path. Because \systemname{} enforces policies on every inter-service interaction, any per-hop overhead is amplified along the critical path.

We report mean, 95th-percentile, and 99th-percentile latency,
since tail latency dominates perceived response time in
multi-hop microservice call paths~\cite{dean_tail_2013}. Measuring tail latency, therefore, allows us to assess how \systemname affects worst-case request behavior under realistic execution conditions.

\begin{figure*}[t]
    \centering
    \begin{subfigure}[t]{0.32\textwidth}
        \centering
        \includegraphics[width=\linewidth]{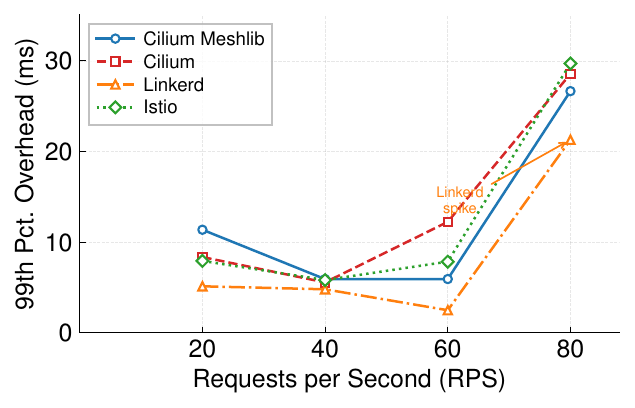}
        \caption{99\%-percentile (ms)}
        \label{fig:result-overhead-99p}
    \end{subfigure}
    \begin{subfigure}[t]{0.32\textwidth}
        \centering
        \includegraphics[width=\linewidth]{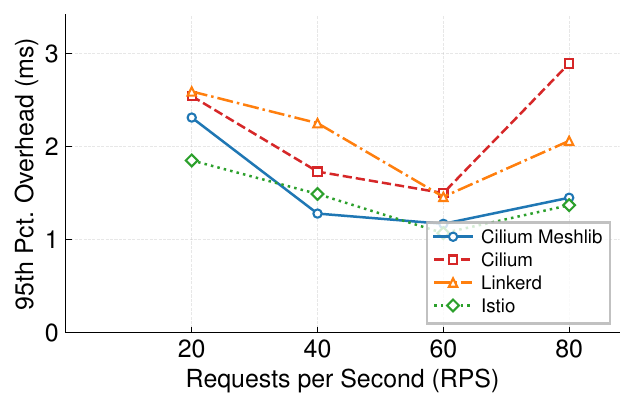}
        \caption{95\%-percentile (ms)}
        \label{fig:result-overhead-95p}
    \end{subfigure}
    \begin{subfigure}[t]{0.32\textwidth}
        \centering
        \includegraphics[width=\linewidth]{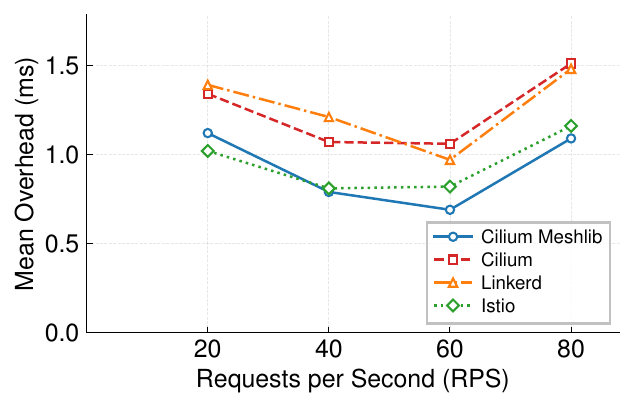}
        \caption{Mean (ms)}
        \label{fig:result-overhead-mean}
    \end{subfigure}
    \caption{Request overhead for benchmark application running on the tested service meshes}
    \label{fig:result-overhead}
\end{figure*}

Figure~\ref{fig:result-overhead} shows the mean, 95th-, and 99th-percentile end-to-end latency for all four configurations. \systemname{} achieves the lowest latency across all three metrics, outperforming Istio, Linkerd, and unmodified Cilium. This outcome is consistent with \systemname{}'s design goal of eliminating the Layer~7 proxy from the critical request path. Both \systemname{} and unmodified Cilium outperform Istio and Linkerd; Cilium's eBPF data plane handles traffic directly in the kernel, avoiding the per-request overhead of a sidecar proxy. The mean latency (Figure~\ref{fig:result-overhead-mean}) shows a smaller gap between \systemname{} and unmodified Cilium than the 99th percentile (Figure~\ref{fig:result-overhead-99p}), suggesting that \systemname{}'s benefit is most pronounced under tail conditions.

Both Cilium with \systemname{} and unmodified Cilium consistently outperform Istio and Linkerd. This difference reflects their use of eBPF-based packet processing in the kernel, which avoids the per-hop overhead introduced by sidecar proxies. Across all configurations, the 99th-percentile latency rises sharply beyond 100~RPS, consistent with database saturation in the TrainTicket benchmark at high load~\cite{zhou_benchmarking_2018}.

\subsection{Request overhead}

To isolate the cost of policy enforcement and control-plane communication, we measure request overhead as the difference between the total client-observed round-trip time and the server-side processing time recorded at the application boundary by the Spring Boot request filter: 

\[  t_{\text{overhead}} = t_{\text{client}} - t_{\text{server}}. \] 

Here, $t_{\text{client}}$ denotes the elapsed time from issuing the request to receiving the response, while $t_{\text{server}}$ excludes network traversal, routing, and policy enforcement. Therefore, their difference precisely quantifies the overhead that \systemname{} introduces to the data path.

\begin{figure}[tp]
    \centering
    \begin{subfigure}[t]{0.48\linewidth}
        \includegraphics[width=\linewidth]{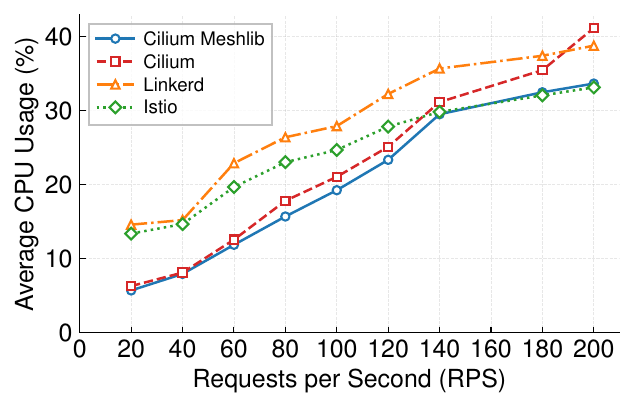}
        \caption{CPU usage (\%)}
        \label{fig:result-cpu}
    \end{subfigure}
    \begin{subfigure}[t]{0.48\linewidth}
        \centering
        \includegraphics[width=\linewidth]{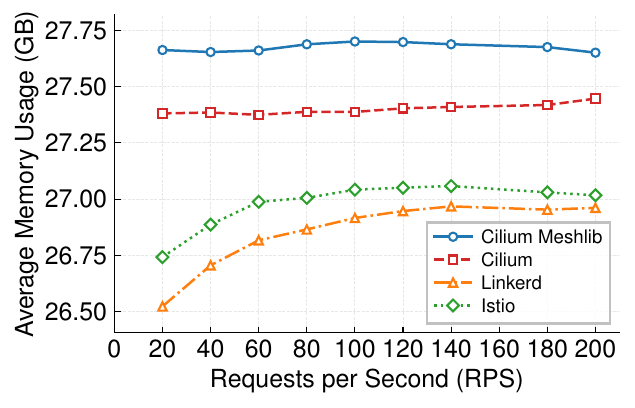}
        \caption{Memory usage in GB}
        \label{fig:result-mb}
    \end{subfigure}
    \caption{Resource consumption of the benchmark application for the tested service meshes}
\end{figure}

Figure~\ref{fig:result-overhead} presents the mean, 95th, and 99th percentiles of request overhead across configurations. Compared to end-to-end latency, overhead measurements exhibit higher variance across request rates. This variability arises from runtime effects such as JVM scheduling and the subtraction-based measurement method, which amplifies noise.

Despite this, \systemname{} does not introduce additional overhead relative to existing service meshes and consistently outperforms unmodified Cilium. However, the absence of a clear trend across all loads suggests that this metric is less stable than end-to-end latency for distinguishing performance differences.

\subsection{Resource consumption}
We also analyze CPU and memory consumption to quantify the resource cost of enforcing \systemname{}'s execution policy. While latency reflects user experience, resource consumption determines the system's efficiency and scalability. Increased CPU or memory usage directly translates into higher deployment costs, particularly in large-scale microservice deployments.

We measure per-service CPU and memory consumption across different configurations to assess how policy enforcement overhead scales with system load and application complexity. These measurements allow us to evaluate the trade-off between the security and isolation guarantees provided by \systemname{} and the operational cost of enforcing them.

\noindent\textbf{CPU usage.}
Figure~\ref{fig:result-cpu} presents the average CPU usage measured during the benchmark runs. Cilium with \systemname and standard Cilium exhibit the lowest CPU usage during low-load scenarios, outperforming non-eBPF-based service meshes. However, during high-load scenarios, this difference becomes much smaller, and Cilium's CPU usage exceeds that of the other service meshes. This could be caused by inefficiencies in the Envoy proxy Cilium uses to parse Layer-7 traffic.

\noindent\textbf{Memory usage.}
Figure~\ref{fig:result-mb} shows the average memory consumption for each service mesh. The results indicate that memory usage differences between the four service mesh configurations are relatively small, with variations on the order of a few hundred megabytes out of a total usage of approximately 27 GB.

Cilium with \systemname has the highest memory consumption among the service meshes tested. This is likely due to the specific benchmark application, TrainTicket, where the application library for \systemname had to be implemented in Java, a memory-intensive programming language.

Given the scale of the overall memory footprint, these differences are unlikely to be significant in practice. Observed differences are small relative to the overall application footprint and are unlikely to materially affect deployment decisions.

\subsection{Evaluation Summary}

The performance evaluation presented in this section provides empirical evidence to answer the research question. First, the results confirm that the primary source of latency overhead in traditional service meshes originates from Layer 7 proxy-based processing. As shown by end-to-end latency measurements, configurations that rely on sidecar proxies, such as Istio and Linkerd, consistently exhibit higher latency than proxy-less approaches. This validates the earlier analysis that redundant parsing, inter-process communication, and additional network hops introduced by proxies significantly contribute to overall latency.

Second, the evaluation shows that architectural modifications to remove the Layer 7 proxy from the critical request path are effective at reducing this overhead. By shifting application-layer policy enforcement into the application itself and relying on Cilium's eBPF-based data plane for lower-layer enforcement, \systemname achieves the lowest end-to-end latency among the evaluated configurations. This confirms that the proposed hybrid architecture successfully reduces latency while remaining compatible with existing service mesh infrastructure.

Third, the experimental results show that these modifications do not degrade the service mesh's functional guarantees. \systemname enforces the same set of security policies as the other evaluated meshes and remains interoperable with services that do not adopt the application library, thereby preserving deployment flexibility and incremental adoptability.

Finally, the resource consumption analysis indicates that the proposed approach does not introduce prohibitive operational costs. CPU usage is comparable to or lower than that of traditional service meshes under most load conditions, and memory overhead remains modest relative to the total application footprint. Together, these findings demonstrate that \systemname answers the central research question by reducing service mesh latency overhead while maintaining security policy enforcement functionality and acceptable resource efficiency.

\newcommand\rot[1]{\rotatebox{90}{%
    \parbox{\widthof{HHHHHHHH\ }}{\centering  #1}}}

\section{Related Work}\label{sec:soa}

Prior work on service mesh design clusters into four
architectural families that differ in where Layer~7
processing executes; Table~\ref{tab:comparison} summarizes their trade-offs.

\begin{table}[t]
\centering
\setlength{\tabcolsep}{2.6pt}
\renewcommand{\arraystretch}{0.9}
\caption{ Comparison of representative service mesh architectures. L7 denotes the Layer~7 processing location. Code indicates whether applications require source-code modifications. Plat.\ denotes tight platform or kernel integration. Latency (Lat.), resource usage (Res.), and flexibility (Flex.) are qualitative relative to a no-mesh baseline. L/M/H denote low, moderate, and high overhead; M-H denotes moderate-to-high overhead. }
\begin{tabular}{lcccccc}
\toprule

System
& L7
& Code
& Plat.
& Lat.
& Res.
& Flex.
\\

\midrule

Istio
& Side.
& \checkmark
& \cross
& H
& H
& H \\

Linkerd
& Side.
& \checkmark
& \cross
& M-H
& M-H
& M \\

Cilium
& --
& \checkmark
& \cross
& L
& L
& L \\

Cilium+Envoy
& Node
& \checkmark
& \cross
& M
& L
& H \\

ServiceRouter~\cite{ServiceRouter-OSDI23}
& App
& \cross
& \cross
& L
& L
& H \\

mRPC~\cite{remote-NSDI23}
& Shared
& \cross
& \checkmark
& L
& L
& L \\

CanalMesh~\cite{song_canal_2024}
& Cent.
& \checkmark
& \cross
& M-H
& L
& M-H \\

KernelOffload~\cite{sidoretti2023application}
& Kernel
& \checkmark
& \checkmark
& L
& L
& M \\

\midrule

\textbf{\systemname}
& \textbf{App}
& \checkmark
& \cross
& \textbf{L}
& \textbf{L}
& \textbf{H} \\

\bottomrule
\end{tabular}
\label{tab:comparison}
\end{table}
\noindent\textbf{Sidecar-based meshes.} Istio and Linkerd deploy a proxy alongside each application instance, intercepting all traffic without requiring code changes~\cite{istio,linkerd}. This provides strong isolation and high configurability, but each request traverses an additional user-space proxy, incurring context switches, data copies, and redundant protocol parsing~\cite{zhu_dissecting_2023}. These costs scale with deployment size, as every service instance requires a dedicated sidecar, and are amplified in high-fan-out call graphs where requests traverse multiple proxies on a single path.
 
\noindent\textbf{Node- and kernel-level approaches.} Cilium moves transport- and network-layer enforcement into the kernel via eBPF, eliminating per-pod proxies for L3/L4 policies~\cite{noauthor_cilium_nodate}. For Layer~7 enforcement, however, it falls back to a node-level Envoy proxy, reintroducing parsing overhead. More aggressive approaches offload L7 logic entirely into the kernel~\cite{sidoretti2023application}, but are constrained by eBPF's verifier restrictions and have not demonstrated the full range of service mesh functionality.
 
\noindent\textbf{Application-integrated designs.} ServiceRouter~\cite{ServiceRouter-OSDI23} and mRPC~\cite{remote-NSDI23} avoid redundant parsing by operating directly on application data structures, achieving low latency and resource consumption. However, they require every service to be modified or recompiled against a proprietary runtime, which limits deployability in heterogeneous clusters and precludes incremental adoption.
 
\noindent\textbf{Centralized proxies.} Canal Mesh~\cite{song_canal_2024} amortizes proxy costs across services by centralizing enforcement into a shared layer, reducing per-instance resource overhead at the cost of additional network hops and weaker per-service isolation guarantees.

\section{Limitations and Future Work}\label{sec:limitations}

\noindent\textbf{Single-node evaluation.} The evaluation is conducted on a single node to isolate enforcement overhead from inter-node network variability. In production environments, deployments typically span multiple nodes, where factors such as cross-rack latency, control-plane scaling under pod churn, and identity-cache update frequency may influence absolute latency values. Nevertheless, \systemname{}'s overhead represents a per-request enforcement cost that remains independent of network topology. Inter-node latency would be added across all configurations and would not alter the relative ordering of results. Extending the evaluation to a multi-node cluster constitutes a primary direction for future research.

\noindent\textbf{Java/Spring Boot library only.} The current \systemname{} library is implemented exclusively for Java/Spring Boot. Services developed in Go, Python, Rust, or other runtimes are unable to utilize the library and instead default to Cilium's Envoy proxy. This limitation is engineering-related rather than inherent to the design, as the gRPC control protocol and enforcement architecture are language-agnostic. Developing libraries for Go and Python, which are the two most prevalent runtimes in production Kubernetes deployments, represents a concrete near-term extension.

\noindent\textbf{Trust model.} In-process enforcement places policy evaluation inside the application process. A compromised application could, in principle, bypass the library's checks. This is a weaker isolation boundary than a sidecar, which enforces policies in a separate process. We note, however, that Cilium's eBPF data plane continues to enforce transport-layer identity independently of the library: a compromised application cannot forge its Cilium security identity or communicate with services it is not permitted to reach at Layer~4. The library adds endpoint-level granularity on top of this baseline; it does not replace it.

\noindent\textbf{HTTP evaluation only.} \systemname{} supports both HTTP and gRPC enforcement, but our evaluation covers HTTP workloads only. gRPC uses protobuf encoding, which has different parsing characteristics from JSON-over-HTTP, and the latency benefit of eliminating redundant parsing may differ. Benchmarking gRPC workloads is left for future work.

The current evaluation utilizes the TrainTicket benchmark~\cite{zhou_benchmarking_2018}, which is the standard workload for service mesh research, but it saturates at approximately 100 RPS due to database contention. Future evaluations should include workloads with higher throughput ceilings and varied service call-graph structures. Policy evaluation is also limited to path and method matching; the performance of header-based and regex-heavy policies has not been characterized.

\section{Conclusion}\label{sec:conc}
This work demonstrates that service mesh latency can be significantly reduced by relocating layer-7 policy enforcement from external proxies to applications. We showed that redundant parsing and user-kernel space transitions in traditional proxies are the primary contributors to latency.
\systemname{}, our hybrid architecture, combines application-integrated policy enforcement with an eBPF-based data plane. By leveraging existing application-level parsing, it eliminates redundant protocol interpretation. Results show lower 99th-percentile latency compared to Istio, Linkerd, and proxy-based Cilium, with particularly significant improvements in multi-microservice deployments. The resource consumption of \systemname remains competitive, with comparable CPU usage and negligible memory overhead.
We showed that fine-grained security enforcement can be application-integrated rather than proxy-centralized, achieving better performance without sacrificing deployability or requiring invasive infrastructure changes.

\balance

\bibliographystyle{ACM-Reference-Format}
\bibliography{paper}

\end{document}